\documentclass[letterpaper, 10 pt, conference]{ieeeconf}  

\IEEEoverridecommandlockouts

\usepackage{graphics} 
\usepackage{epsfig} 
\usepackage{mathptmx} 
\usepackage{graphicx}
\usepackage{float}

\usepackage{url}

\usepackage{breakurl}
\usepackage[breaklinks]{hyperref}

\usepackage{censor}

\title{\LARGE \bf
Establishing Human-Robot Trust through Music-Driven Robotic Emotion Prosody and Gesture
}

\author{Richard Savery, Ryan Rose and Gil Weinberg$^{1}$
\thanks{$^{1}$Georgia Tech Center for Music Technology, Atlanta, GA, USA
        {\tt\small rsavery3@gatech.edu}}%
}

\begin{document}

\maketitle
\thispagestyle{empty}
\pagestyle{empty}

\begin{abstract}
As human-robot collaboration opportunities continue to expand, trust becomes ever more important for full engagement and utilization of robots. Affective trust, built on emotional relationship and interpersonal bonds is particularly critical as it is more resilient to mistakes and increases the willingness to collaborate. In this paper we present a novel model built on music-driven emotional prosody and gestures that encourages the perception of a robotic identity, designed to avoid uncanny valley.
Symbolic musical phrases were generated and tagged with emotional information by human musicians. These phrases controlled a synthesis engine playing back pre-rendered audio samples generated through interpolation of phonemes and electronic instruments. Gestures were also driven by the symbolic phrases, encoding the emotion from the musical phrase to low degree-of-freedom movements. Through a user study we showed that our system was able to accurately portray a range of emotions to the user. We also showed with a significant result that our non-linguistic audio generation achieved an 8\% higher mean of average trust than using a state-of-the-art text-to-speech system.

\end{abstract}

\section{INTRODUCTION}
 As co-robots become prevalent at home, work, and in public environments, a need arises for the development of trust between humans and robots. A meta-study of human-robot trust \cite{hancock2011meta} has shown that robot-related attributes are the main contributors to building trust in Human-Robot-Interaction, affecting trust more than environmental and human related factors. Related research on artificial agents and personality traits \cite{Mateas:1999:ORI:1805750.1805762,Bates:1994:REB:176789.176803} indicates conveying emotions using subtle non-verbal communication channels such as prosody and gesture is an effective approach for building trust with artificial agents. These channels can help convey intentions as well as expressions such as humor, sarcasm, irony, and state-of-mind, which help build social relationship and trust. 
 
 In this work we developed new modules for Shimi, a personal robotic platform \cite{bretan2015emotionally}, to study whether emotion-driven non-verbal prosody and body gesture can help establish affective-based trust in HRI. Our approach is to use music, one of the most emotive human experiences, to drive a novel system for emotional prosody \cite{adolphs2001emotion} and body gesture \cite{shan2007beyond} generation. We propose that music-driven prosody and gesture generation can provide effective low degrees of freedom (DoF) interaction that can convey robotic emotional content, avoid the uncanny valley \cite{macdorman2006subjective}, and help build human-robot trust. Furthermore, trust is highly dictated by the first impression for both human-human and human-robot relations \cite{8525669}, implying methods for gaining trust at the start of a relationship, such as prosody and gesture are crucial.

\begin{figure}[h]
  \centering
\includegraphics[angle=270,width=7cm]{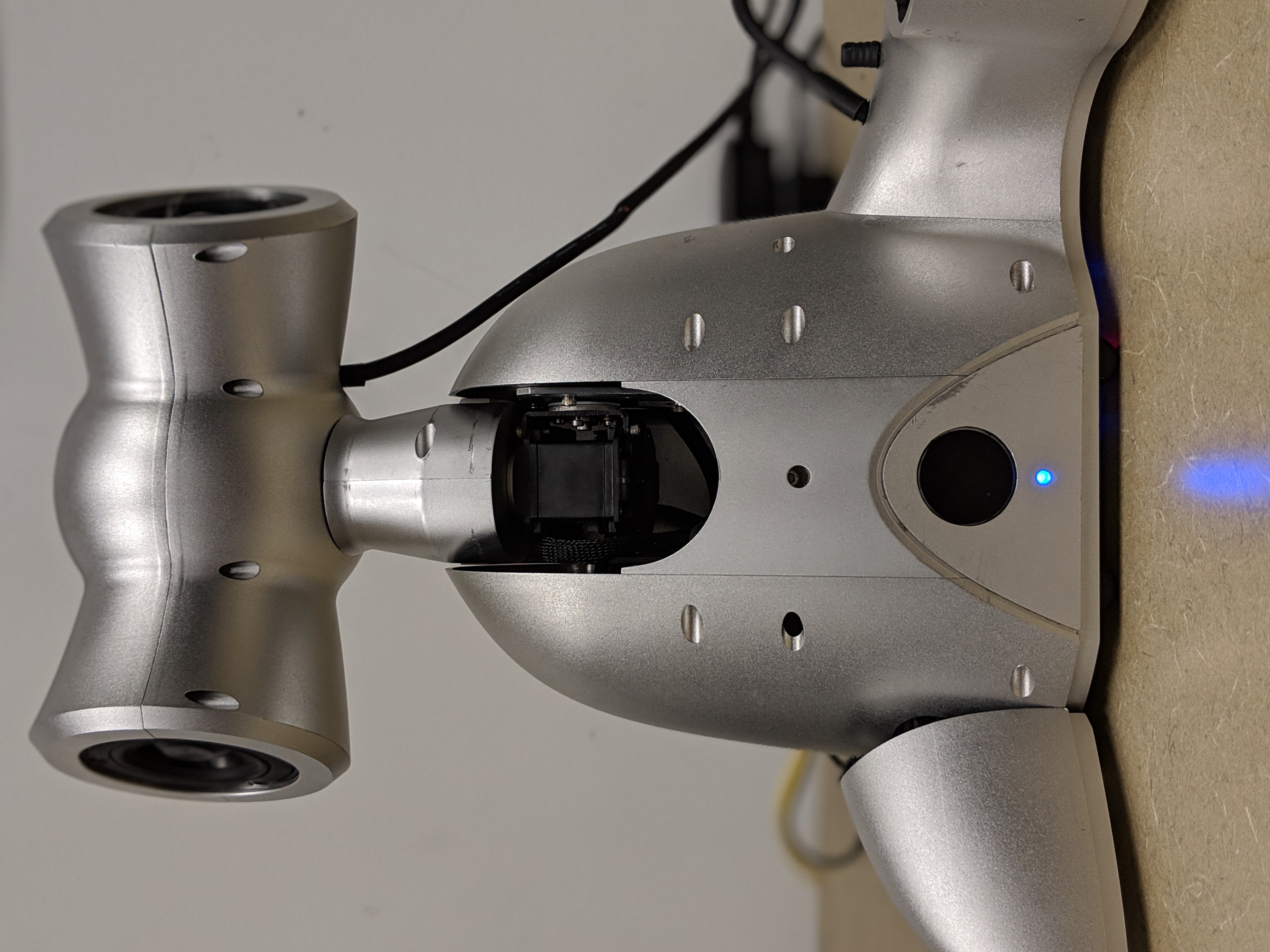}
  \caption{The musical robot companion Shimi}
  \label{fig:shimi}
\end{figure}

We address two research questions, firstly, can we use non-verbal prosody through musical phrases combined with gestures to accurately convey emotions? We present a novel deep learning based musical voice generation system that uses the created phrases to generate gesture through musical prosody. In this way music is central to all interactions presented by Shimi. We evaluate the effectiveness of the musical audio and generative gesture system for Shimi to convey emotion, specified by valence-arousal quadrants. Our second research question is whether emotional conveyance through prosody and gestures driven by music analysis can increase the level of trust in human-robot-interaction. For this we conduct a user study to evaluate prosodic audio and gestures created by our new model in comparison to a baseline text-to-speech system.

\section{Background}
\subsection{Trust in HRI}
Trust is a key requirement for working with collaborative robots, as low levels of trust can lead to under-utilization in work and home environments \cite{johndlee}. A key component of the dynamic nature of trust is created in the first phase of a relationship \cite{kim2009repair,miles1995organizational}, while lack of early trust building can remove the opportunity for trust to develop later on \cite{Schaefer2016}. Lack of trust in robotic systems can also lead to expert operators bypassing the robot to complete tasks \cite{satchell2016cockpit}.
Trust is generally categorized into either \textit{cognitive trust or affective trust} \cite{freedy2007measurement}. Affective trust involves emotional bonds and personal relationships, while cognitive trust focuses on considerations around dependability and competence. Perceiving emotion is  crucial for the development of affective trust in human-to-human interaction \cite{rousseau1998not}, as it increases the willingness to collaborate and expand resources bought to the interactions \cite{gompei2018factors}.  Importantly, relationships based on affective trust are more resilient to mistakes by either party \cite{rousseau1998not}, and perceiving an emotional identity has been shown to be an important contributor for creating believable and trustworthy interaction \cite{Mateas:1999:ORI:1805750.1805762,Bates:1994:REB:176789.176803}. In group interactions, emotional contagion - where emotion is spread between a group - has been shown to improve cooperation and trust in team exercises \cite{barsade2002ripple}. 

\subsection{Emotion, Music and Prosody}
 Emotion conveyance is one of the key elements for creating believable agents \cite{Mateas:1999:ORI:1805750.1805762}, and prosody has been proven to be an effective communication channel to convey such emotions for humans \cite{wang2013study} and robots \cite{crumpton2016survey}.  On a related front, music which shares many of the underlying building blocks of prosody such as pitch, timing, loudness, intonation, and timbre \cite{wennerstrom2001music}, has also been shown to be a powerful medium to convey emotions \cite{sloboda1999music}. 
 In both music and prosody, emotions can be classified in a discrete categorical manner (happiness, sadness, fear, etc.) \cite{devillers2005challenges}, and through continuous dimensions such as valence, arousal, and less commonly, dominance, and stance \cite{russell2009emotion, mehrabian1996pleasure}. While some recent efforts to generate and manipulate robotic emotions through  prosody focused on linguistic robotic communication \cite{crumpton2016survey}, \cite{breazeal2002recognition} no known efforts have been made to use models from music analysis to inform real-time robotic non linguistic prosody for collaboration, as we propose here.

\subsection{Emotion, Music and Gesture}
 Human bodily movements are embedded with emotional expression \cite{InderbitzinVCVB11,walbott98}, which can be processed by human  ``affective channels" \cite{degelderNature}. Researchers in the field of affective computing, have been working on designing  machines that can process and communicate emotions through such channels \cite{Picard95affectivecomputing}. Non-conscious affective channels have been demonstrated to communicate compassion, awareness, accuracy, and competency, making them vital components of social interaction with robots \cite{scheutz2006utility}. Studies have shown clear correlations between musical features and movement features, suggesting that a single model can be used to express emotion through both music and movement \cite{sievers_music_2013,weinberg2009zoozbeat,weinberg2006jam}. Certain characteristics of robotic motion have been shown to influence human emotional response \cite{riek2010cooperative,moon2013design}. Additionally, emotional intelligence in robots leads to facilitated human-robot interactions \cite{kozima2001search,breazeal2002recognition,castellano2010affect}. Gesture has been used to accompany speech in robots to help convey affective information \cite{alibali2000gesture}, however, to our knowledge, there is no prior work that attempts to integrate physical gestures and music-driven prosody to convey robotic emotional states. 

\section{Shimi and Emotion}
\subsection{Prosody}
The goal of this project was to create a new voice for Shimi, using musical audio phrases tagged with an emotion. We aimed to develop a new voice that could generate phrases in real-time. This was achieved through a multi-layer system, combining symbolic phrase generation using MIDI controlling a synthesis playback system. 

\begin{figure}[h]
  \centering
\includegraphics[width=8cm]{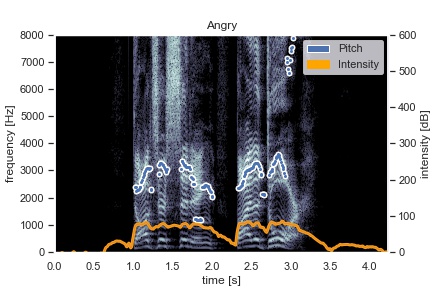}
  \caption{Angry Speech Pitch and Intensity}
  \label{fig:angry}
\end{figure}

\begin{figure}[h]
  \centering
\includegraphics[width=8cm]{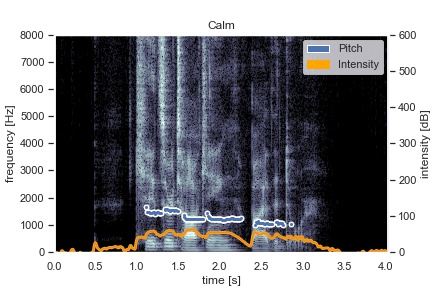}
  \caption{Calm Speech Pitch and Intensity}
  \label{fig:calm}
\end{figure}

\subsubsection{Dataset and Phrase Generation}
\label{sec:technical_description_dataset}
To control Shimi's vocalizations we generate MIDI phrases that drive the synthesis and audio generation described below and lead the gesture generation. MIDI is a standard music protocol, where notes are stored by pitch value with a note on velocity, followed by a note off to mark the end of the note. With the absence of appropriate datasets we chose to create our own set of MIDI files tagged with valence and arousal by quadrant. MIDI files were collected from eleven different improvisers around the United States, each of whom tagged their recorded files with an emotion corresponding to a quadrant of the valence/arousal model. Phrases were required to be between 100ms and 6 seconds and each improviser recorded between 50 to 200 samples for each quadrant. To validate this data we created a separate process whereby the pitch range, velocities and contour were compared to the RAVDESS \cite{ravdess} data set, with files removed when the variation was over a manually set threshold. RAVDESS contains speech files tagged with emotion, Figure \ref{fig:angry} and Figure \ref{fig:calm} clearly demonstrate the variety of prosody details apparent in the RAVDESS dataset (created using \cite{Parselmouth,praat}) and the variation between a calm and angry utterance of the same phrase.


We chose to use a Recurrent Neural Network, Long Short Term Memory (RNN-LSTM) as described in {\cite{deepscore}} to generate musical phrases using this data. RNN-LSTM's have been used effectively to generate short melodies, as they are sequential and consider the input as the output is generated. Other network structures were considered however we found RNN-LSTM's very effective for this task when compared to alternate approaches developed by the authors \cite{bretan2016unit,savery2018interactive,savery2015algorithmic,saverysoccer}, more detail is provided in \cite{savery_finding_2019}. 
While using samples directly from the data was possible, using a neural network allowed infinite variation but also allowed for generated phrases utilizing all the musical features created by the improvisers, not just one improviser per sample.

\subsubsection{Audio Creation and Synthesis}
\label{sec:technical_description_dataset}
The generated MIDI phrases contain symbolic information only without audio. To create audio we developed a synthesis system to playback the MIDI phrase. As we desired to create a system devoid of semantic meaning a new vocabulary was constructed. This was built upon phonemes from the Australian Aboriginal language Yuwaalaraay a dialect of the Gamilaraay language. Sounds were created by interpolating four different synthesizer sounds with 28 phonemes. Interpolation was done using a modified version of WaveNet. These samples are then time stretched and pitch shifted to match the incoming MIDI file. For a more detailed technical overview of audio processing read \cite{savery_finding_2019}. 

\subsection{Gestures}
In human communication gestures are tightly coupled with speech \cite{mcneill2012language}. Thus, Shimi's body language is implemented in the same way, derived from its musical prosody and leveraging the musical encoding of emotion to express that emotion physically. Music and movement are correlated, with research finding commonalities in features between both modes \cite{sievers_music_2013}. Additionally, humans demonstrate patterns in movement that is induced from music \cite{toiviainen_embodied_2010}. Particular music-induced movement features are also correlated to perceived emotion in music \cite{burger_relationships_2013}. After a musical phrase is generated for Shimi's voice to sing, the MIDI representation of that phrase is provided as input to a gesture generation system. Musical features such as tempo, range, note contour, key, and rhythmic density are obtained  from the MIDI through Python libraries \verb|pretty_midi| \cite{raffel} and \verb|music21|\footnote{https://github.com/cuthbertLab/music21}. These features are used to create mappings between Shimi's voice and movement: for example, pitch contour is used to govern Shimi's torso forward and backward movement. Other mappings include beat synchronization across multiple subdivisions of the beat in Shimi's foot, and note onset-based movements in Shimi's up-and-down neck movement.

After mapping musical features to low-level movements, Shimi's emotional state is used to condition the actuation of the movements. Continuous values for valence and arousal are used to influence the range, speed, and amount of motion Shimi exhibits. Some conditioning examples include limiting or expanding the range of motion according to the arousal value, and governing how smooth motor direction changes are through Shimi's current valence level. In some cases, the gestures generated for one degree of freedom are dependent on another degree of freedom. For example, when Shimi's torso leans forward, Shimi's attached head will be affected as well. As such, to control where Shimi is looking, any neck gestures need to know the position of the torso. To accommodate these inter-dependencies, when the gesture system is given input, each degree of freedom's movements are generated sequentially and in full, before being actuated together in time with Shimi's voice. Video examples of gesture and audio are available at www.richardsavery.com/shimitrust.

\section{Methodology}
We designed an experiment to identify how well participants could recognize the emotions shown by our music-driven prosodic and gestural emotion generator. This part of the experiment aimed to answer our first research question, can non-verbal prosody combined with gestures accurately portray emotion. After watching a collection of stimuli, participants completed a survey measuring the trust rating from each participant. This part of the experiment was designed to answer the second question, can emotion driven, non-semantic audio generate trust in a robot.

We hypothesized that through non-semantic prosodic vocalizations accompanied with low-DoF robotic gesture humans will be able to correctly classify Shimi's portrayed emotion as either happy, calm, sad, or angry, with an accuracy consistent with that of text-to-speech. Our second hypothesis was that we will see higher levels of trust from the Shimi using non-speech.

\subsection{Stimuli}

\begin{table}[h]
\begin{center}
\begin{tabular}{||c c c c||} 
\hline
  Name  & Audio  & Stochastic & Experimental \\
 \hline
 Audio Only& X & &    \\ 
 \hline
 Stochastic Gesture, audio & X & X & \\ 
 \hline
 Stochastic Gesture, no audio &  & X & \\
 \hline
 Experimental Gesture, audio & X &  & X \\
 \hline
  Experimental Gesture,&  & X & X \\
  no audio &  &  &  \\
 \hline
\end{tabular}
\caption{Experiment Stimuli} \label{tab:stimuli} 
\end{center}
\end{table}
\begin{figure*}[h]
  \centering
\includegraphics[width=12cm]{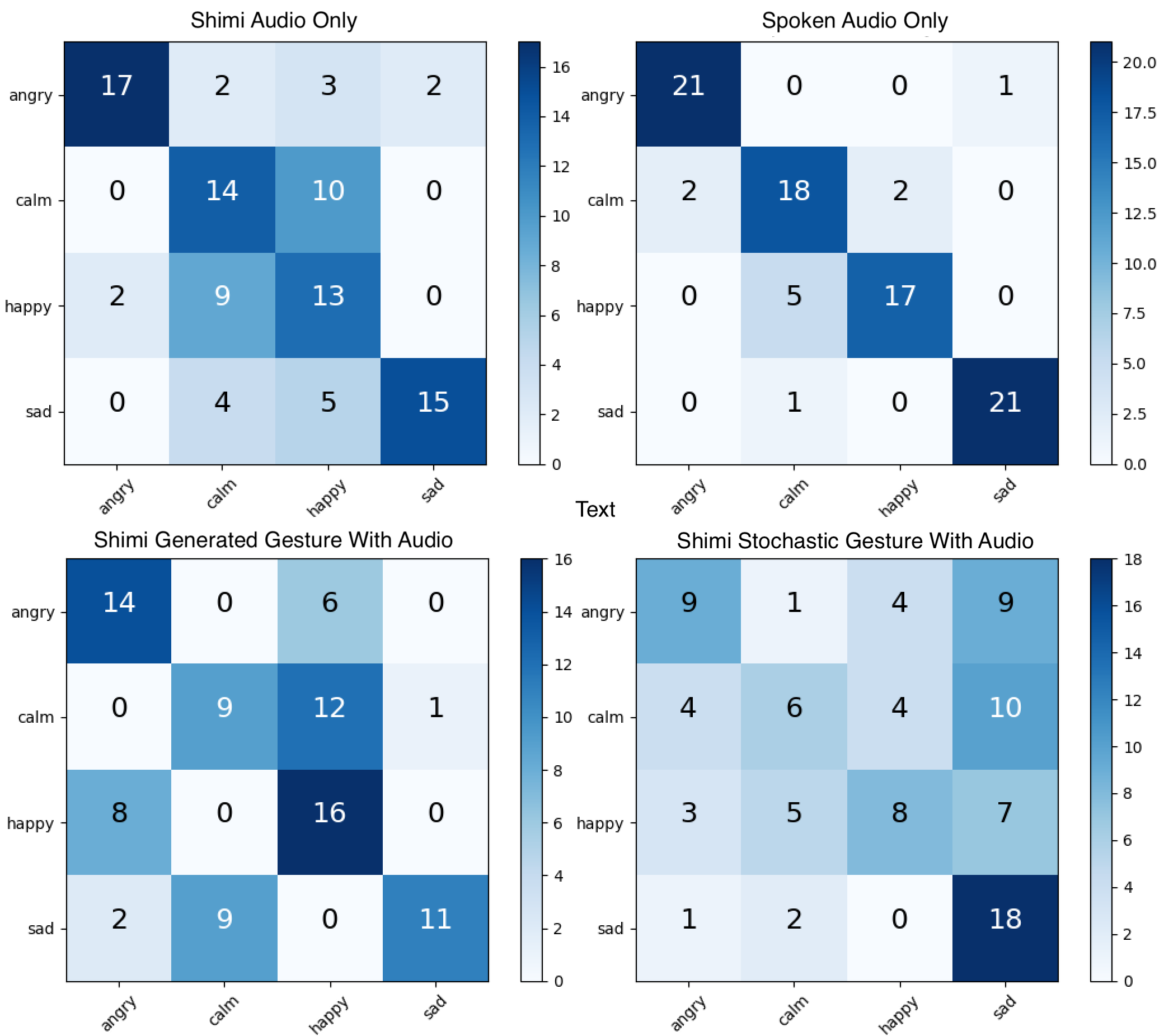}
  \caption{Confusion Matrix}
  \label{fig:matrix}
\end{figure*}
The experiment was designed as a between-subjects study, where one group would hear the audio with the Shimi voice, while the other would hear pre-rendered text-to-speech. Both groups saw the same gesture and answered the same prompts. The text-to-speech examples were synchronized in length and emotion to Shimi's voice. The stimuli for the Speech Audio experiment used CereProc's Meghan voice\footnote{https://www.cereproc.com/}. CereProc is a state of the art text to speech engine. The text spoken by Meghan was chosen from the EmoInt Dataset \cite{MohammadB17wassa}, which is a collection of manually tagged tweets. 

\subsection{Emotion}
The generated gestures were either deterministic gestures created using the previously described system, or deterministic stochastic gestures. Stochastic gestures were implemented by considering each DoF separately, restricting their ranges to those implemented in the generative system, and specifying random individual movement durations up to half of the length of the full gesture. The random number generator used in these gestures were seeded with an identifier unique to the stimuli such that they were deterministic between participants. Gesture stimuli were presented both with and without audio. 

\subsection{Procedure}
Participants were gathered from the undergraduate student population at the Georgia Institute of Technology (N=24). Subjects participated independently, with the group alternating for each participant, culminating with 12 in each group. The session began with an introduction to the task of identifying the emotion displayed by Shimi. Participants responded through a web interface that controlled Shimi through the experiment and then allowed the user to select the emotion they thought Shimi was expressing. Stimuli were randomly ordered for each participant. Table \ref{tab:stimuli} shows the order of stimuli used, each category contained 8 stimuli, 2 for each valence arousal quadrant. After identifying all stimuli participants were directed to a Qualtrics survey to gather their trust rating.

To measure trust, we used the Trust Perception Scale-HRI \cite{Schaefer2016}. This scale uses 40 questions, each one using a rating scale between 0-100\%, to give an average trust rating per participant. The questions take between 5-10 minutes to complete and include questions such as how often the robot will be reliable or pleasant. After completing the trust rating, participants had several open text boxes to discuss any observations in regards to emotion recognition, trust or the general experiment. This was the first time trust was mentioned in the experiment. 

\section{Results}
 \begin{figure*}[h]
  \centering
\includegraphics[width=18cm]{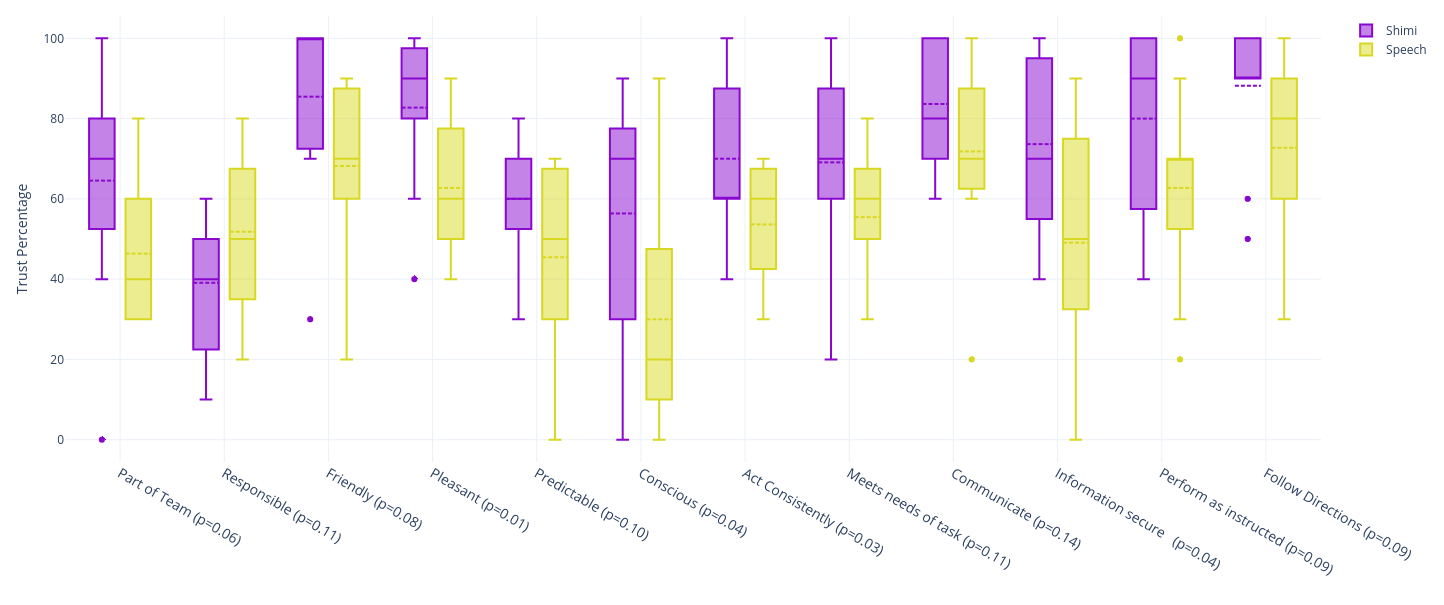}
  \caption{Questions with P less than 0.1}
  \label{fig:shimisig}
\end{figure*}

\subsection{Gestures and Emotion}
After data was collected, two participant's emotion prediction data was found to be corrupted due to a problem with the testing interface, reducing the number of participants in this portion of the study to 22. First, we considered classification statistics for the isolated predictions of Shimi's voice and text-to-speech (TTS) voice. While TTS outperformed Shimi's voice (F1 score $TTS=0.87$ vs. $Shimi=0.63$), the confusion matrices show errant predictions in similar scenarios (see figure \ref{fig:matrix}). For example, both audio classes struggle to disambiguate happy and calm emotions.

Our hope was that adding gestures to accompany the audio would help to disambiguate emotions. To test that our gestures properly encoded emotion, we compared predictions for Shimi's voice accompanied by generated gestures with predictions accompanied by stochastic gestures, the results of which can also be seen in figure \ref{fig:matrix}.

While the confusion matrices show a clear prediction improvement in using generated gestures over stochastic, the results are not statistically significant. A two-sided T-test provides a p-value of 0.089, which does not reject the null hypothesis at $\alpha=0.05$. Disambiguities from the audio-only cases were not mitigated, but the confused emotions changed slightly, following other gesture and emotion studies \cite{lim_towards_2012}.

\begin{figure}[h]
  \centering
\includegraphics[width=9cm]{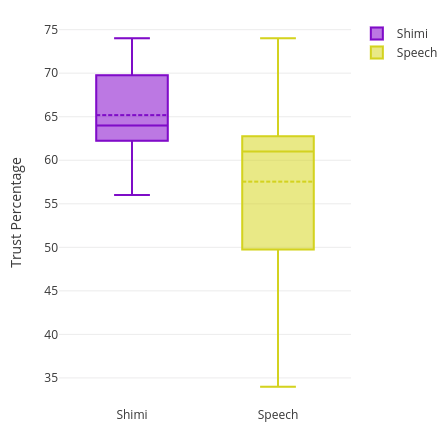}
  \caption{Participants Trust Mean}
  \label{fig:trustmean}
\end{figure}

Some experimental error may have accrued through the mixing of stimuli when presented to participants. Each stimuli was expected to be independent but some verbal user feedback expressed otherwise, such as: ``the gestures with no audio seemed to be frequently followed by the same gesture with audio, and it was much easier to determine emotion with the presence of audio." The presentation of stimuli may have led participants to choose an emotion based on how we ordered stimuli, rather than their perceived emotion of Shimi.

\subsection{Trust}
As per the trust scale, a mean percentage for trust was calculated on combined answers to 40 questions from each participant. A t-test was then run on each group mean. The average score variation between speech and Shimi audio showed a significant result (p=0.047), proving the hypothesis. Figure \ref{fig:trustmean} shows the variation in average scores from all participants. The difference of mean between groups was 8\%. 
Results from the text entries were positive for the prosodic voice, and generally neutral or often blank for speech. A common comment from the participants for the Shimi voice was ``Seemed like a trustworthy friend that I would be fine confiding in."

\begin{figure}[h]
  \centering
\includegraphics[width=9cm]{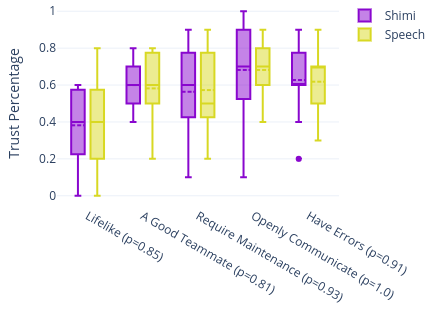}
  \caption{Not Significant Trust Results}
  \label{fig:notsig}
\end{figure}

\section{Discussion and Future Work}
We were able to clearly demonstrate participant recognition of the expected emotion from Shimi, confirming our first hypothesis. Our model however did not perform completely as predicted, as audio without gesture lead to the clearest display of emotion. With a small sample size and a p-value close to being significant, we were encouraged by qualitative feedback that provided insight into the shortcomings of the gestures and gave us ideas for future improvements. For instance, emotions on the same side of the arousal axis were often hard to disambiguate. One participant noted that ``it was generally difficult to distinguish happy and angry if there were no sounds (similar situation between sad and calm)", while another noted ``I had some trouble discerning calm from sad here and there", and ``without speaking, it was difficult to decipher between anger and excitement".  The general intensity of the emotion was apparent, however." Certain movement features led to emotional connections for the participants, as demonstrated here: ``generally, when Shimi put it's head down, I was inclined to say it looked sad. When it moved more violently, particularly by tapping [sic] it's foot, I was inclined to say it was angry or happy", ``more forceful movements tended to suggest anger", and ``When there was more severe motion, I associated that with anger. When the motion was slower I associated it with sad or calm. If the head was down more I associated it with sad. And I associated it with happy more when there was sound and more motion."

The trust perception scale is designed to give an overall rating, and independent questions should not necessarily be used to draw conclusions. However, there were several interesting results indicating further areas of research.  Fig \ref{fig:shimisig} shows all categories with a p value less  than 0.10, for which multiple questions showed significant results with a p value under 0.05). Shimi's voice was crafted to be friendly and inviting and as expected received much higher results for pleasantness and friendliness.  Unexpectedly, it also showed much higher ratings for its perception as being conscious. While further research is required to confirm the meaning, we believe that the question on consciousness of Shimi demonstrating a significant result shows that embodying a robot with a personal prosody (as opposed to human speech) creates a more believable agent. Figure \ref{fig:notsig} shows the categories with very similar distributions of scores. These include Lifelike, A Good Teammate, Have Errors, Require Maintenance and Openly Communicate. While further research is needed, this may imply that these features are not primarily associated with audio. Further work should be done to explore if the same impact can be found by adjusting audio features of a humanoid robot may also lead to interesting results. 

In other future work we plan to develop experiments with a broader custom musical data-set across multiple robots. We intend to study emotional contagion and trust between larger groups of robots across distributed networks \cite{weinberg1999expressive}, aiming to understand collaboration and trust at a higher level between multiple robots.

Overall, our trust results were significant and showed that prosody and gesture can be used to generate higher levels of trust in human-robot interaction. Our belief that creating a believable agent that avoided uncanny valley was shown to be correct and was validated through participant comments, including the open text response: ``Shimi seems very personable and expressive, which helps with trust".


\bibliographystyle{IEEEtran}
\bibliography{name}









\end{document}